\documentclass[a4paper,12pt]{article}
\usepackage[latin2]{inputenc}
\usepackage[T1]{fontenc}
\usepackage{graphicx}

\title{Network analysis of correlation strength between the most developed countries.}
\author{Janusz Miśkiewicz\\
  Institute of Theoretical Physics, University of Wroc\l{}aw, \\
pl. M. Borna 9, 50-204 Wroc\l{}aw, Poland \\
Department of Physics and Biophysics, \\
Wroc\l{}aw University of Environmental and Life Sciences \\
ul. Norwida 25, 50-375 Wroc\l{}aw}

\begin{document}

\maketitle

\begin{abstract}
A new algorithm of the analysis of correlation among economy time series is proposed. The algorithm is based on the power law classification scheme (PLCS) followed by the analysis of the network on the percolation threshold (NPT). The algorithm was applied to the analysis of correlations among GDP per capita time series of 19 most developed countries in the periods (1982, 2011), (1992, 2011) and (2002, 2011). The representative countries with respect to strength of correlation, convergence of time series and stability of correlation  are distinguished. The results are compared with ultrametric distance matrix analysed by NPT.

\end{abstract}

\section{Introduction}

Economy system is a perfect example of a complex system were various factors interact. Investigation of those interactions is a non-trivial task. On the other hand knowledge about relationship between economy entities is crucial while preparing strategy and taking decisions.  Analysis of correlations among significant number of entities  can be divided into two steps: correlation analysis and correlation matrix analysis. The second part arises due to the fact that the correlation matrix consists of hundreds (or thousands of elements) and it would be difficult to analyse each element separately. Since the series of works by Mantegna and collaborators \cite{Mantegna1999,Mantegna1999b,Mantegna1999c,Bonanno2004,Bonanno2001,Bonanno2000,DiMatteo2004} the standard approach to tackle the correlation problem in an economy system is to calculate distance matrix between entities and based on it build a minimum spanning tree (MST), which provide various information about clusterings, sectors, leaders/hubs etc. The 
analysis is in 
fact an extension of Markowitz portfolio analysis \cite{Markowitz1952}, where Pearson correlation coefficient plays a key role in portfolio optimisation procedure. 
Although, very successful in stock market analysis in the case of macroeconomy systems the other algorithms (with different distance metrics or network structures used) seam to be more useful \cite{Ausloos2010,Janusz2010,Janusz2008,Miskiewicz2012,Miskiewicz2009,Miskiewicz2010,Miskiewicz2008a,bib:jm_rozdzial}. The main disadvantages of the Mantegna algorithm results from  the choice of the distance metrics. It have been shown  \cite{Miskiewicz2012,bib:jm_rozdzial} that the ultrametric distance (UD) is sensitive to the presence of noise and wrongly classify nonlinear correlations. Therefore it is very disputable whenever UD reflects fully the relationship between time series. Moreover, the second part of the Mantegna algorithm i.e. MST construction is in fact a very strong restriction influencing structure of generated network. 
Recall that the MST algorithm solves the problem to find a graph which connects all the vertices, but is the smallest considering the number of links (or lowest sum of the links weights) \cite{graham1985history,frieze1985value}. Within the present paper an alternative algorithm of the time series correlation analysis is proposed. The algorithm consists of two elements: (i) correlation matrix construction, where instead of the UD a power law classification scheme (PLCS) is used and (ii) the analysis of the graph on the percolation threshold.

\section{Time series correlation analysis}
\subsection{Power law classification scheme}

The standard correlation analysis is based on two fundamental methods: Pearson correlation coefficient \cite{Rodgers1988} and regression analysis e.g. \cite{Sobczyk2006}. Both methods are focused on linear correlation investigations. Recently an interesting detrended cross-correlation analysis (DDCA) was proposed \cite{Podobnik2008}, which allows to investigate long-range correlations with power law cross-correlation function.  Sometimes ``back-doors'' are used to apply regression analysis to investigate nonlinear correlations by applying appropriate transformations of variable i.e. log-log transformation makes possible to apply regression analysis to verify power law dependencies. However, if the nonlinear dependency of unknown form is expected various rank methods, entropy based methods or numerous hypothesis tests \cite{Wackerly2008,Sharma2005,Tsay2010,Mirkin2011,Kantz2003,Redelico2010,Redelico2009} are used. Unfortunately they are able to answer the question if the correlation exists, but cannot find the 
character of the correlation. Therefore, there is a need to develop a method which could investigate nonlinear dependencies between time series. In the present paper the Manhattan Distance (MD) \
\cite{bib:taxicab,Krause1987} (known also as taxi cab distance) will be used as a starting point of the proposed correlation analysis method. MD is the cumulative sum of the absolute value of the differences between two time series and, as it have been shown \cite{bib:jm_fens2010,bib:jm_rozdzial}, it is more robust to the noise presence than Pearson correlation coefficient besides allows to analyse nonlinear dependencies. The crucial feature of MD is the fact that its value depends on the length of the time series -- it is a nondecreasing function of its length. 

Let consider two time series $A$ and $B$ and denotes theirs elements as $a_i$ and $b_i$ respectively. The correlation between the time series will be defined as:
\begin{equation}
 a_i=f(b_i),
\label{eq:correlation}
\end{equation}
where $i \in (1, \ldots , N)$. $N$ -- is the number of the elements in considered time series A or B. It is assumed that the time series are of equal length or only the overlapping parts are considered.
MD distance between $A$ and $B$ is defined:
\begin{equation}
 M\!D(A,B) = \sum_{i=1}^{N} |a_i-b_i|
\label{eq:MD}
\end{equation}
As it was mentioned MD is a nondecreasing function of the time series length therefore the time series A and B can be transformed into cumulative time series of MD:
\begin{equation}
 M\!D(A,B)(j) = \sum_{i=1}^{j} |a_i-b_i|, j\in (1,2,\ldots , N)
\label{eq:MD_cum}
\end{equation}
The cumulative time series $M\!D(A,B)(j)$ can be replaced by a continuous function $M\!D(A,B)(x)$, where $x \in (1,N)$. Moreover, the cumulative time series $M\!D(A,B)(j)$ is strictly related to the surface between time series A and B. The procedure is in fact similar to the definition of the Riemann integral or various procedures of numerical integration e.g. \cite{Isaacson1994,Woodford2011}.

Let us approximate time series A and B by a continuous function and in order to avoid absolute value notation assume that $a_i \leq b_i$, additionally let assume that time series A and B are correlated Eq.(\ref{eq:correlation}), then
\begin{equation}
M\!D (A,B)(n) \simeq \int_0^n (a(t) - f(a(t))) \, dt. 
\end{equation}
Considering that the integral is calculated along A time series the correlation function can be found by:
\begin{equation}
f(n)= \frac{d M\!D (A,B)(n)}{dn}.
\label{eq:diff_corr}
\end{equation}

In principal the method is capable to detect arbitrary correlations, but due to the practical constrains such as a limited length of time series, stationarity of time series, presence of noise, etc. it is useful to distinguish a class of correlation functions. In analogy to the algorithm classification complexity \cite{Homenda2009,Papadimitriou2002} a class of correlation function is introduced. Since one of the main aims of the method is to measure ``strength'' of correlation the power laws functions are chosen as a basis of the method. Moreover, from the practical point of view power law functions are relatively easy to fit. Additionally, they are frequently reported in the analysis of financial data e.g. \cite{Todorova20114433,PhysRevE.84.046112,0295-5075-95-6-68002}.

Considering the above the power law classification scheme (PLCS) is defined as follows:
\begin{enumerate}
\item Transform time series A and B into a cumulative MD time series (Eq.(\ref{eq:MD_cum})).
\item Present the calculated cumulative MD time series as a function of the time series length in logarithmic scale.
\item Fit the linear function to the data and find the slope coefficient $\alpha$ corresponding to the integrated correlation function.
\item Let the class of the correlation function be labelled as $\gamma (A,B)=\alpha -1$ (see Eq.(\ref{eq:diff_corr})).
\item Additionally, calculate the quality of the fit, measured by the statistical significance probability $\beta$. It provides information about the stability of the correlation in the considered time interval.
\end{enumerate}

\subsection{Ultrametric Distance}

The ultrametric distance (UD) \cite{Mantegna1999c} can be considered as a standard in analysis of relationship among shares \cite{Skornik-Pokarowska2004,Naylor2007,Mizuno2006,Brida2010} therefore is used to compare with the results of PLCS. For the convenience of the reader the main definitions and basic features of UD will be recalled here.
The Pearson correlation coefficient is defined as:
\begin{equation}
\label{eq:correlations}
P_{(t,T)} (A,B) = \frac{\langle A B \rangle_{(t,T)} - \langle A \rangle_{(t,T)} \langle B \rangle_{(t,T)} }{\sqrt{( \langle A^2 \rangle_{(t,T)} - \langle A \rangle^2_{(t,T)}) ( \langle B^2 \rangle_{(t,T)} - \langle B \rangle^2_{(t,T)}})}.
\end{equation}
The UD in literature is used in two forms:
\begin{equation}
\label{eq:stat_M}
DU_{MS}(A,B)_{(t,T)} = \sqrt{2(1- P_{(t,T)} (A,B))},
\end{equation}
as introduced by Mantegna \cite{Mantegna1999c} or
\begin{equation}
\label{eq:stat_AM}
DU_{AM}(A,B)_{(t,T)} = \sqrt{\frac{1}{2}(1- P_{(t,T)} (A,B))},
\end{equation}
introduced in \cite{Miskiewicz2008a}. The difference between Eq.(\ref{eq:stat_M}) and Eq.(\ref{eq:stat_AM}) lies in normalization. The Mantegna formulation map distance between time series into the interval $(0,2)$, while the definition Eq.(\ref{eq:stat_AM}) maps the distance between time series into the interval $(0,1)$, where zero corresponds to the linearly correlated time series, one refers to anticorrelated time series and $\frac{\sqrt{2}}{2}$ to the case where linear correlation was not found. In the following Eq.(\ref{eq:stat_AM}) will be used and denoted by UD.

\subsection{Network analysis}

The distance or, in general, correlation analysis in the case of investigations of group of stocks, bonds, shares or economy indices results in distance/correlation matrix, which consists of hundreds of data. The detailed analysis of such a matrix is beyond standard investigations. Moreover, usually researches are not interested in the correlation itself but rather what kind of structure of dependencies do they form. Therefore, the correlation/distance matrix is usually analysed by constructing and analysing a chosen network structure. In the case of correlation analysis based on UD the most popular structure is the minimum spanning tree (MST). It should be noticed that the network structure can influence strongly the results of the analysis. Therefore, in the present analysis a network on the percolation threshold is chosen. This network is constructed by removing links from the smallest (or strongest) correlations. Links are removed until the next removed link leads to breakdown of the network into 
disconnected subnetworks. Finally, the structure of network on percolation threshold (NPT) is investigated. 

\section{Correlations of developed countries}

The analysis of correlations was performed for the group of developed countries (in alphabetic order): Australia (AU), 	Austria (AT),	Belgium (BE),  Canada (CA),	 Denmark (DN),	France (FR),	Germany (DE), Great Britain (GB), Greece (GR),	Holland (NL),  Italy (IT), 	Ireland (IR), 	Japan (JP),	Luxembourg (LU),	Poland (PL), 	Portugal (PT), Spain (SE),	Sweden (SW),	 USA (US).
The countries were characterised by the GDP per capita time series. The time series were downloaded from the Mathematica data base and covered the interval (1970, 2011). The correlation analysis was conducted for the short (10 years: (2002, 2011)), medium (20 years: (1992, 2011)) and long (30 years: (1982, 2011)) time span. PLCS algorithm is based on MD therefore correlation strength and stability commutes i.e. $\gamma (A,B) = \gamma (B,A)$ and $\beta (A,B) = \beta (B,A)$. Therefore, the correlation  and stability matrix consists of 171 unique values.

\subsection{Key results of PLCS}

The detailed analysis of a correlation matrix is a laborious process, therefore it is analysed  by constructing the chosen network structure. However, for the sake of clarity some representative cases will be discussed in details.

{\bf The negative correlation strength. }

\begin{figure}
 \centering
 \includegraphics[scale=0.6]{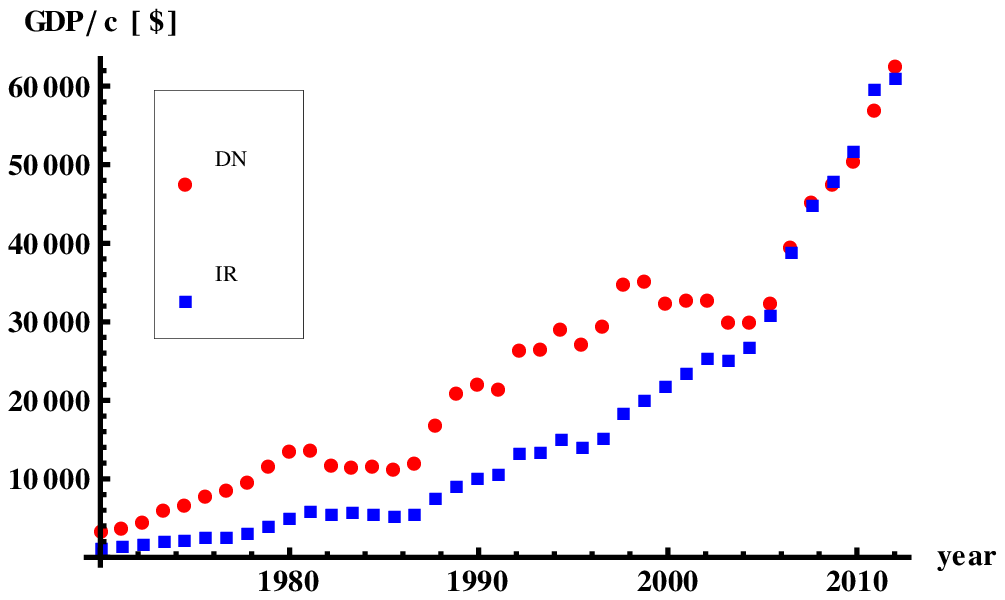}
 \includegraphics[scale=0.6]{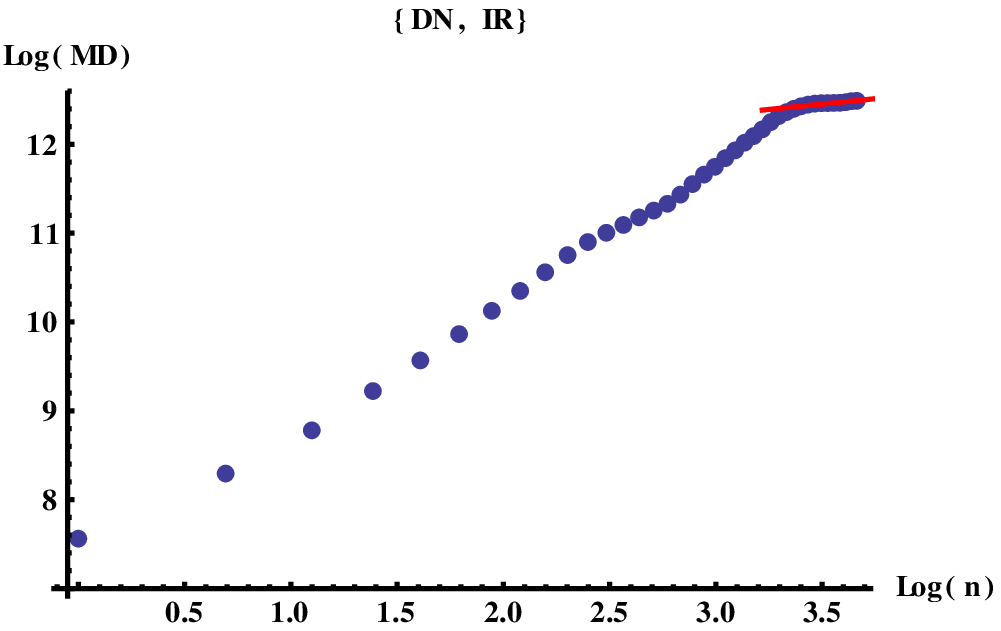}
 \caption{Left: The GDP per capita of Denmark and Ireland in US [\$] in years (1970, 2011). Right: Cumulative MD between Denmark and Ireland as a function of the time series length in logarithmic scale. A linear function (red line) is fitted to ten data points at large $n$.}
 \label{fig:DN_IR_10Y}
\end{figure}

The negative correlation strength $\gamma$ corresponds to the situation where MD between analysed time series is decreasing so the time series are converging. The example of negative correlation strength are presented in Tab.\ref{tab:PLCS_nagative}.  The plots of GDP/c and cumulative MD in logarithmic scale with fitted linear function to the data points at large $n$ are presented in the case of the countries with the lowest correlation strength -- Denmark and Ireland (Fig.\ref{fig:DN_IR_10Y}). The PLCS negative correlation strength corresponds to the evolution of the GDP/c of DN and IR. In fact since 1998 GDP/c of those countries are converging and since 2005 they are overlapping. The UD of pairs presented in Tab.\ref{tab:PLCS_nagative} takes values in the range (0.05; 0.19) which might be considered as a indication of possible linear correlation.

\newcommand{\mc}[3]{\multicolumn{#1}{#2}{#3}}

\begin{table}
\begin{center}
\begin{tabular}{|c|l|l|l|}
\hline
\mc{4}{|c|}{10 years (2002, 2011)} \\ \hline
\mc{1}{|c|}{pair} & \mc{1}{c|}{$\gamma$}&\mc{1}{c|}{$\beta$} &\mc{1}{c|}{UD} \\ \hline
 (DN, IR) & -0.76 & $ 4.3 \cdot 10^{-5}$ & 0.10   \\ \hline
 (SW, GB) & -0.53 & $ 1.2 \cdot 10^{-6}$ & 0.10 \\ \hline
(NL, SW) & -0.53& $ 1.9 \cdot 10^{-8}$ & 0.06 \\ \hline
 \mc{4}{|c|}{20 years (1992, 2011)} \\ \hline
(DN, IR) & -0.76 & $ 1.0 \cdot 10^{-6}$ & 0.14   \\ \hline
(SW, GB) &-0.53 & $ 1.7 \cdot 10^{-4}$ & 0.19   \\ \hline
(NL, SW) & -0.53 & $ 3.3 \cdot 10^{-9}$ & 0.14   \\ \hline
 \mc{4}{|c|}{30 years (1982, 2011)} \\ \hline
(IT, AU) & -0.18 & $ 8.4 \cdot 10^{-21}$ & 0.16 \\ \hline
(AT, DE) & -0.16 & $ 1.0 \cdot 10^{-11}$ & 0.05  \\ \hline
 (IT, CA) & -0.13 & $  4.0 \cdot 10^{-20}$ & 0.14  \\ \hline

\end{tabular}
\end{center}
\caption{Correlation strength correlation stability  and UD of the first three pairs of countries with the most negative $\gamma$ in recent ten (2002, 2011), twenty (1992, 2011) and thirty years (1982, 2011).}
\label{tab:PLCS_nagative}
\end{table}

{\bf The correlation strength close to zero and one. }

\begin{figure}
 \centering
 \includegraphics[scale=0.6]{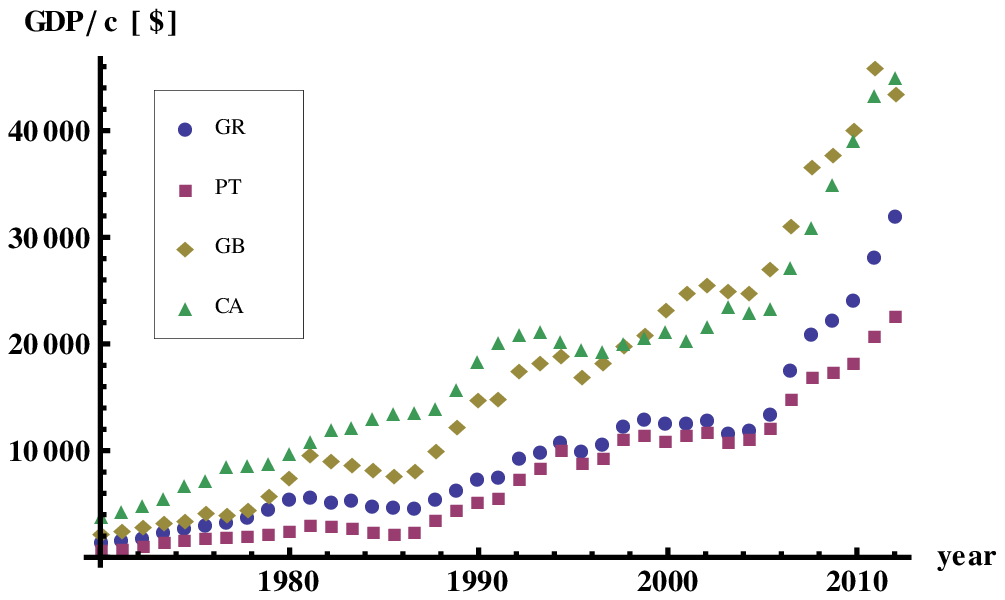}
 \includegraphics[scale=0.6]{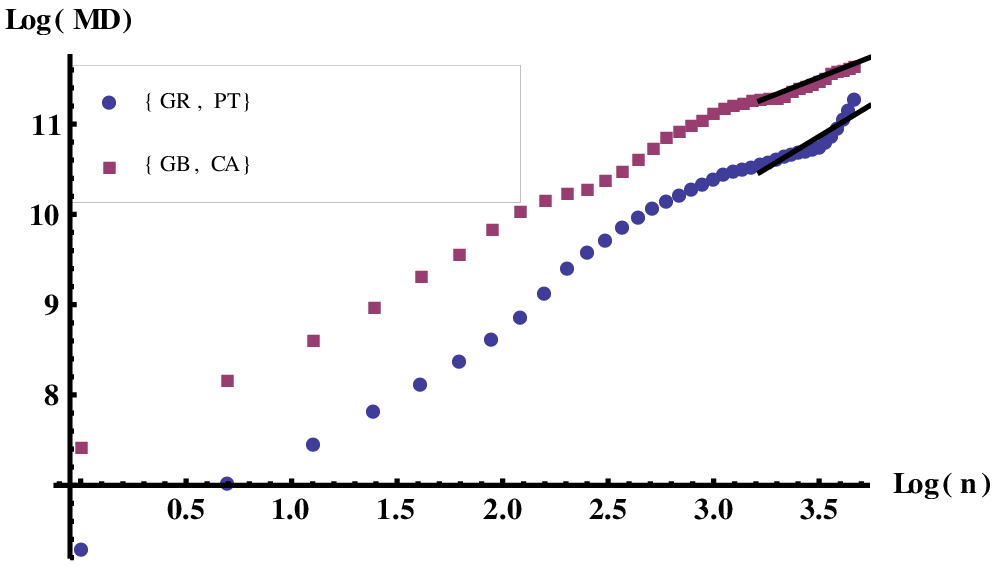}
 \caption{Left: The GDP per capita of Greece, Portugal, Great Britain and Canada in US [\$] in years (1970, 2011). Right:  Cumulative MD between (Greece, Portugal) and (Great Britain, Canada) as a function of the time series length in logarithmic scale. A linear functions (black lines) are fitted to ten data points at large $n$ for every pair.}
 \label{fig:zero_one}
\end{figure}

The correlation strength equal zero or one corresponds to the situation where the time series are linearly correlated. The first one, where
correlation strength is close to zero, is in fact a trivial correlation where the time series are overlapping. The pairs for which the correlation strength was close to zero are presented in Tab.\ref{tab:PLCS_zero}. 
The plot examples of cumulative MD with the fitted linear functions in the case of correlation strength is close to zero or one ((GB, CA) and (GR, PT) respectively) are presented in Fig.\ref{fig:zero_one}. 
In such a case UD might classify the dependence as a linear correlation, although in the case of correlation between (GR, PT) the results might changed if different time interval is chosen. This example illustrate the possibility of two trends which can compensate. The results of PLCS algorithm and UD are particularly equivocal for GDP/c time series of Belgium and Austria for the recent 30 years, where $\gamma =-0.003$ and UD=0.04. On the other hand very interesting is the case two pairs Great Britain -- Canada and Great Britain -- USA. These countries are known from the close mutual cooperation. The correlation strength $\gamma \sim 0$ indicates linear correlation of those time series, however UD value is greater then zero (UD(GB, CA)= 0.10, UD(GB, US)= 0.12), which can be also interpreted as a presence of some noise.  Moreover, this supposition is supported by low stability of these correlations, which is equal $3.3 \cdot 10^{-9}$ and $1.5 \cdot 10^{-6}$ respectively. Notice that $\beta$ of correlation 
strength for the considered 171 pairs of countries in the period (2002, 2011) is varying from $2.2 \cdot 10^{-19} $ to $3.2 \cdot 10^{-4}$. Comparing the stability of correlation for the pairs 
(GB, CA) and (GB, US) 
and the results or the whole group it must be concluded that $\beta$ lies close to low stability cases.  

The second case of the correlation strength equal one is the linear correlation. It is very interesting that PLCS point out clear, linear correlation between GDP/c of Greece and Portugal, particularly in the view of recent crisis in Greece and difficult situation in Portugal. On the other hand linear correlation is observed between Poland and Belgium with high stability of the correlation ($\beta = 3.5 \cdot 10^{-11}$), which suggests that there is a chance for a good prognosis for Poland.

 \begin{table}
\begin{center}
\begin{tabular}{|c|l|l|l|c|l|l|l|}
\hline
\mc{8}{|c|}{10 years (2002, 2011)} \\ \hline
\mc{1}{|c|}{pair} & \mc{1}{c|}{$\gamma$}&\mc{1}{c|}{$\beta$} &\mc{1}{c|}{UD}& \mc{1}{|c|}{pair} & \mc{1}{c|}{$\gamma$}&\mc{1}{c|}{$\beta$} &\mc{1}{c|}{UD} \\ \hline
(GB, CA) & -0.009 &$3.3 \cdot 10^{-9}$ & 0.10 & (GR, PT) & 1.0 & $2.3 \cdot 10^{-5}$ & 0.03 \\ \hline
 (GB, US) & 0.028 & $1.5 \cdot 10^{-6}$ & 0.12 & (BE, US) & 1.0 & $1.4 \cdot 10^{-6}$ & 0.16 \\ \hline
 \mc{8}{|c|}{20 years (1992, 2011)} \\ \hline
(SW, GB) & -0.014 & $1.6 \cdot 10^{-11}$ &0.19 & (BE, PL) & 1.0 & $3.5 \cdot 10^{-11}$ & 0.15  \\ \hline
 (IR, CA) & 0.008 &$ 3.4 \cdot 10^{-11}$ & 0.14 & (DE, AU) & 1.0 & $3.9 \cdot 10^{-9}$ & 0.19 \\ \hline 
 \mc{8}{|c|}{30 years (1982, 2011)} \\ \hline
(BE, AT) & -0.003 & $ 3.1 \cdot 10^{-27}$ & 0.04 & (IT, PT) & 1.0 &  $2.2 \cdot 10^{-35}$ & 0.10 \\ \hline
(NL, GB) & 0.007 & $9.3 \cdot 10^{-26}$ & 0.09 & (LU, PT) & 1.0 &  $1.1 \cdot 10^{-28}$ & 0.08 \\ \hline
\end{tabular}
\end{center}
\caption{Correlation strength, correlation stability and UD of the pairs of countries with the linearly correlated GDP/c time series in recent ten (2002, 2011), twenty (1992, 2011) and thirty years (1982, 2011).}
\label{tab:PLCS_zero}
\end{table}

{\bf High value of correlation strength.}

\begin{figure}
 \centering
 \includegraphics[scale=0.6]{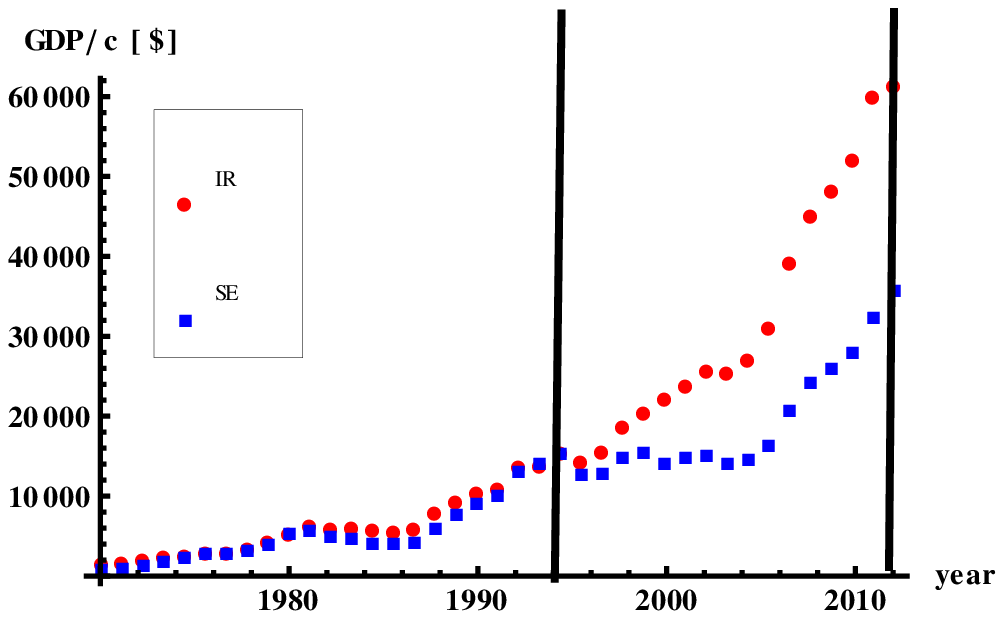}
\includegraphics[scale=0.6]{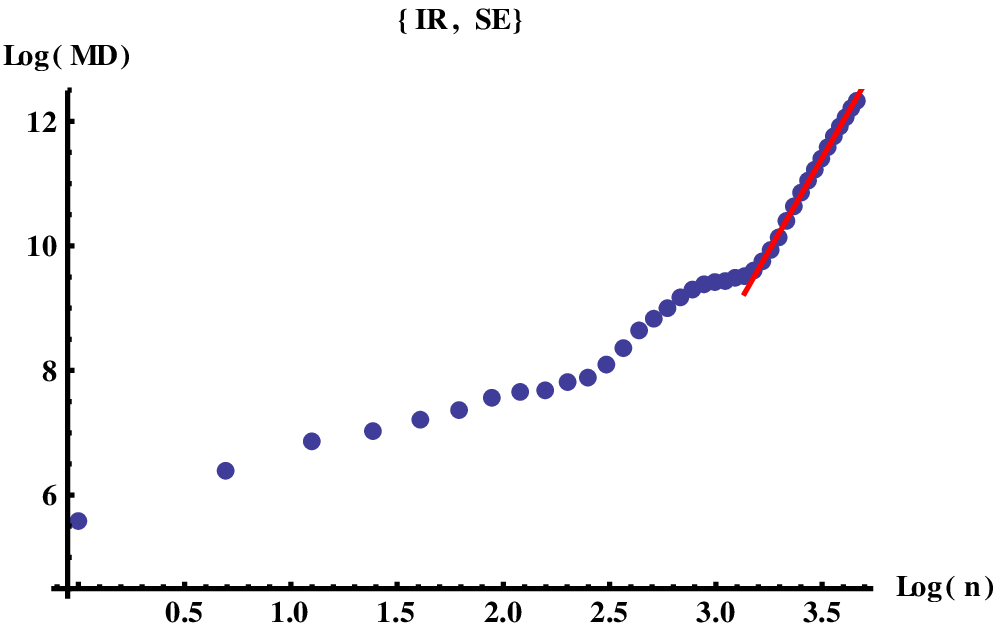}
 \caption{IR SE 10Y
Left: The GDP per capita of Ireland and Sweden in US [\$] in years (1970, 2011). The vertical lines denote the interval for which the linear function was fitted.  Right:  Cumulative MD between Ireland and Sweden as a function of the time series length in logarithmic scale. A linear function (red line) is fitted to fifteen data points at large $n$.}
 \label{fig:IR_SE}
\end{figure}

The positive value of correlation strength indicates that the analysed time series are divergent. The highest strength of correlation was observed in the case of Ireland and Spain. The GDP/c time series and cumulative MD time series in logarithmic scale are presented in Fig.\ref{fig:IR_SE}. The fit of linear function was extended to the interval (1995, 2011), which corresponds to the divergent part of GDP/c time series (Fig.\ref{fig:IR_SE} left). It is worth noticing that UD of this pair of countries is equal 0.07 (Tab.\ref{tab:PLCS_high}), which might suggest existence of linear correlation between these time series. Comparing PLCS and UD results it must be concluded that PLCS gives more adequate description of the system. Of course, the detailed analysis should include analysis of various economy/political/social factors which is beyond the scope of this paper. However, it is worth noticing that correlation understood in a broader then only linear sense takes place in this case. Let take a closer look in 
the 
GDP/c time series of  Ireland and Spain (Fig.\ref{fig:IR_SE}) in the interval (1995, 2011). Two stages of 
development can be distinguished in this period: (i) (1995, 2004) and (ii) (2004, 2011), while in the second subinterval the GDP/c of both countries are growing rapidly. Despite the change in 2004 the correlation remains stable and the cumulative MD is growing according to the well established power law ($\beta = 2.5 \cdot 10^{-15}$). 

\begin{table}
\begin{center}
\begin{tabular}{|c|l|l|l|}
\hline
\mc{4}{|c|}{10 years (2002, 2011)} \\ \hline
\mc{1}{|c|}{pair} & \mc{1}{c|}{$\gamma$}&\mc{1}{c|}{$\beta$} &\mc{1}{c|}{UD} \\ \hline
 (IR, SE) & 4.8 &$ 2.5 \cdot 10^{-15}$ & 0.07 \\ \hline
 (GR, IR) & 3.7 &$ 4.4\cdot 10^{-17}$ & 0.08 \\ \hline 
(FR, NL) & 3.4 &$ 6.7 \cdot 10^{-13}$ &0.04  \\ \hline
 \mc{4}{|c|}{20 years (1992, 2011)} \\ \hline
(LU, SW) & 4.0 & $ 5.4 \cdot 10^{-11 }$ &0.14  \\ \hline
(IR, SE) & 3.6 & $ 2.5 \cdot 10^{-15 }$ &0.13  \\ \hline
(GR, IR) & 3.1 & $ 4.4  \cdot 10^{-17 }$ &0.11  \\ \hline
 \mc{4}{|c|}{30 years (1982, 2011)} \\ \hline
(GR, IR) & 2.7 & $ 7.3 \cdot 10^{-30}$ & 0.09  \\ \hline
(LU, SW) &2.4 & $ 2.2 \cdot 10^{-17}$ & 0.12  \\ \hline
(IR, SE) & 2.2 & $  1.3 \cdot 10^{-18}$ & 0.12  \\ \hline
\end{tabular}
\end{center}
\caption{The correlation strength, stability and UD of the last three pairs of countries with the highest $\gamma$ in recent ten (2002, 2011), twenty (1992, 2011) and thirty years (1982, 2011).}
\label{tab:PLCS_high}
\end{table}

{\bf Stability of correlation}

  \begin{figure}
 \centering
 \includegraphics[scale=0.6]{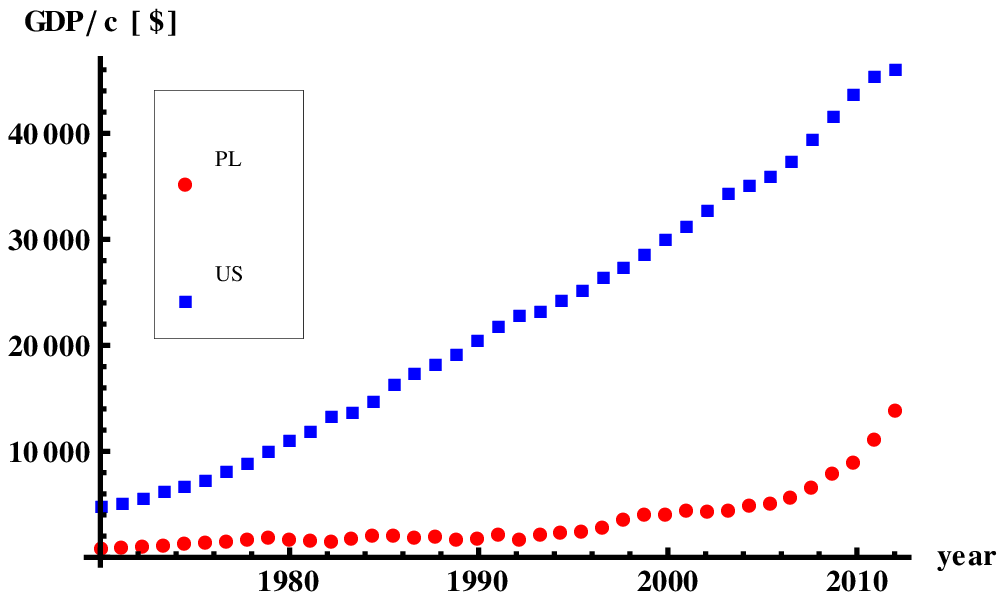}
 \includegraphics[scale=0.6]{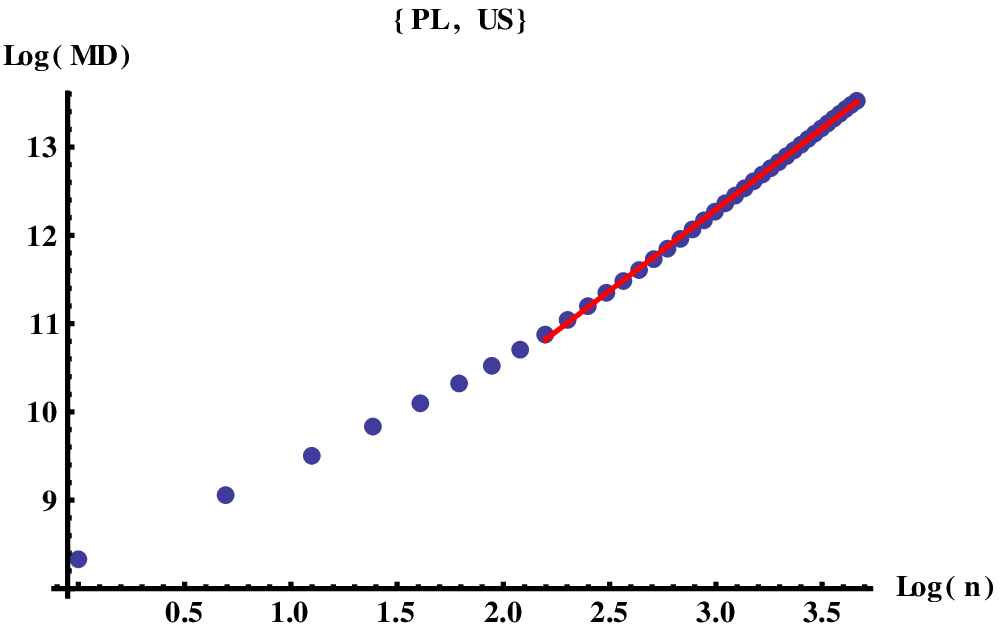}
 \caption{The high stability of correlation on the example of  Poland and USA. Left: GDP/c time series of Poland and USA in US [\$] in years (1970, 2011). Right: Cumulative MD between Poland and USA as a function of the time series length in logarithmic scale. A linear function (red line) is fitted to thirty data points at large $n$. }
 \label{fig:PL_US}
\end{figure}

The PLCS algorithm besides correlation strength provides also the information about stability of the correlation given by a significance level -- the probability at which the null hypothesis can be rejected. The examples of the three most and less stable pairs are presented in Tab.\ref{tab:PLCS_stability}. The very interesting case is the pair Poland -- USA. This pair is present in the first three pairs with the most stable correlation in all of the considered intervals. The GDP/c and the cumulative MD with fitted linear function of (PL, US) are presented in Fig.\ref{fig:PL_US}. The UD of the pair (PL, USA) is equal 0.18, 0.19, 0.24 for the periods ten, twenty and thirty years respectively. The UD results are far from the linear correlation and it is very disputable whenever they can be interpreted as any correlation. PLCS, contrary to UD points out that nonlinear correlation exists and place the correlation among the most stable correlations. This case illustrate one more PLCS features. In the analysis of 
correlation 
between time series of significantly different values global trend is more important than fluctuations. Application of MD in PLCS  places small fluctuations  at the right position and cover them by the global trend. This perfectly agree with first impression of the GDP/c graph (Fig.\ref{fig:PL_US}).

\begin{table}
\begin{center}
\begin{tabular}{|c|l|l|l||c|l|l|l|}
\hline
\mc{4}{|c||}{High correlation stability} & \mc{4}{c|}{Low correlation stability} \\ \hline
\mc{8}{|c|}{10 years (2002, 2011)} \\ \hline
\mc{1}{|c|}{pair} & \mc{1}{c|}{$\gamma$}&\mc{1}{c|}{$\beta$} &\mc{1}{c||}{UD}& \mc{1}{|c|}{pair} & \mc{1}{c|}{$\gamma$}&\mc{1}{c|}{$\beta$} &\mc{1}{c|}{UD} \\ \hline
(PL, US) & 0.9 &$ 2.2 \cdot 10^{-19 } $ & 0.18 & (BE, DE) & 0.2 & $8.1 \cdot 10^{-5 } $ & 0.04  \\ \hline
(DN, AU) & 1.1  &$ 1.5 \cdot 10^{-17 } $ & 0.04 & (IT, AU) &0.3  & $ 1.7 \cdot 10^{-4 } $ &  0.06 \\ \hline
(GR, IR) & 3.7  &$ 4.4 \cdot 10^{- 17} $ & 0.08 & (NL, GB) &-0.5  & $3.2 \cdot 10^{-4 } $ &  0.20 \\ \hline
 \mc{8}{|c|}{20 years (1992, 2011)} \\ \hline
(PL, US) &1.9  &$2.2 \cdot 10^{-19 } $ & 0.19 & (BE, DE) & 1.2 & $ 8.2 \cdot 10^{-5 } $ & 0.09  \\ \hline
(DN, AU) & 2.1  &$ .5 \cdot 10^{-17 } $ & 0.10 & (IT, AU) & 1.3 & $ 1.7 \cdot 10^{-4 } $ &  0.12  \\ \hline
(GR, IR) & 4.7  &$ 4.4 \cdot 10^{-17 } $ & 0.11 & (NL, GB) & 0.5  & $ 3.3 \cdot 10^{-4 } $ & 0.13  \\ \hline
 \mc{8}{|c|}{30 years (1982, 2011)} \\ \hline
(PT, US) & 0.7 &$ 1.5 \cdot 10^{-50 } $ & 0.12 & (FR, NL) & 0.8 & $ 1.8 \cdot 10^{-16 } $ & 0.07   \\ \hline
(PL, US) & 0.8 &$ 5.5 \cdot 10^{-49 } $ & 0.24 & (LU, JP) & 0.6 & $ 1.0 \cdot 10^{-14 } $ &  0.33 \\ \hline
(GR, US) & 0.8 &$ 9.1 \cdot 10^{-47 } $ & 0.19 & (AT, DE) & -0.2 & $ 1.0 \cdot 10^{-11 } $ &  0.05 \\ \hline
\end{tabular}
\end{center}
\caption{The correlation strength, stability and UD of the pairs of countries with the highest and the lowest correlation stability in recent ten (2002, 2011), twenty (1992, 2011) and thirty years (1982, 2011). }
\label{tab:PLCS_stability}
\end{table}

\section{Structure of the networks on the percolation threshold}

The analysis of the correlation matrix is performed by construction of the networks on the percolation threshold (NPT). 
\begin{enumerate}
 \item The algorithm of the NPT begins with sorting the links between countries.
 \item In the next step of NPT algorithm last link of the chosen sorting is removed.
 \item The integrity of the graph is verified.
 \item If the graph is connected the second step of the algorithm is repeated. \\
If the graph is not connected then the recently removed link is added the graph and the algorithm finished.
\end{enumerate}

The main advantage of the network analysis is the ability to identify particularly interesting nodes. In economics, such points are leaders, which influence strongly other, sensitive entities, which reflecting processes in related countries and well established relationships. 
The desired knowledge can be gained through by construction NPT with appropriate sorting. Considering the PLCS results there are three types of useful sorting: from the smallest correlation strength to the biggest, from the biggest correlation strength to the smallest and from the lowest value of correlation strength to the biggest. The first sorting corresponds to the situation where converging time series are preferred, the second sorting is particularly useful when strong non-linear relationship is suspected for which even small variations in one time series results in strong fluctuation in the second. The last sorting prefers the most stable correlations, so the most probable. Therefore it can be constructed three possible networks: convergent (NPT-CP), strength (NPT-SP) and stability preferential (NPT-S).

\begin{figure}
 \centering
 \includegraphics[scale=0.8]{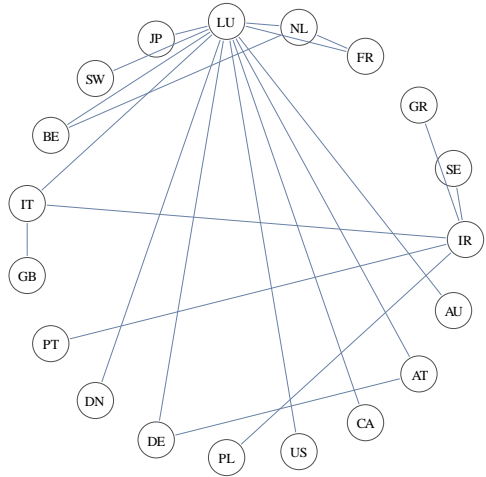}
 \caption{Convergent preferential NPT constructed on the PLCS correlation strength matrix for the time period (2002, 2011). }
 \label{fig:PLCS_rev_str_NPT}
\end{figure}

\begin{figure}
 \centering
 \includegraphics[scale=0.8]{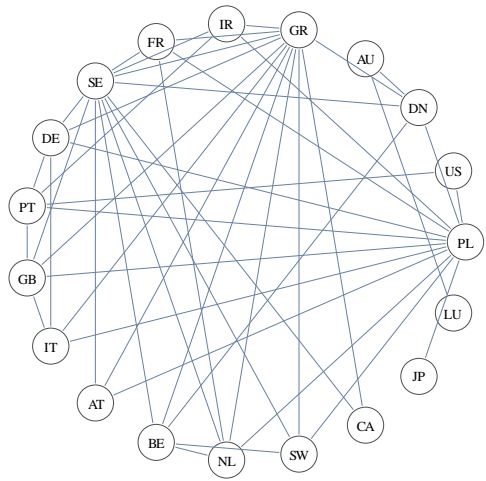}
 \caption{Stability preferential NPT based on the PLCS correlation stability matrix  for GDP/c time series of chosen countries in years (2002, 2011). }
 \label{fig:PLCS_str_NPT}
\end{figure}

\begin{figure}
 \centering
 \includegraphics[scale=0.8]{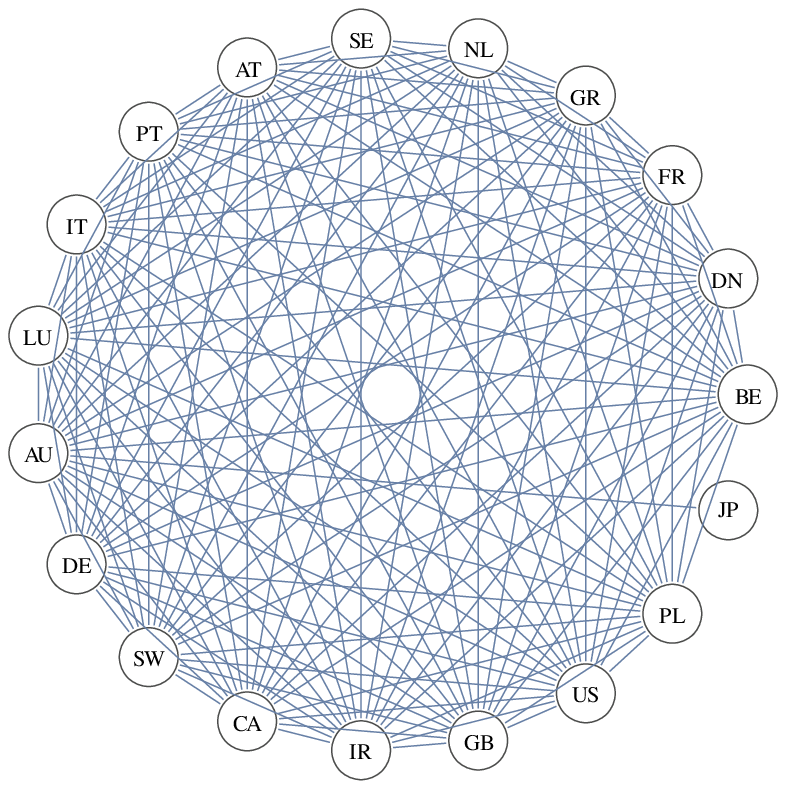}
 \caption{NPT based on the UD matrix for GDP/c time series of chosen countries in years (2002, 2011). }
 \label{fig:UD_NPT}
\end{figure}

The examples of NPT based PLCS analysis of GDP/c time series in the time interval (2002, 2011) are presented in Figs.\ref{fig:PLCS_rev_str_NPT}-\ref{fig:UD_NPT}.  Due to the limitation of the space graphs of NPT constructed for twenty and thirty-year periods are available from the author on request.
The main hub of NPT-CP network (Fig.\ref{fig:PLCS_rev_str_NPT}) is Luxembourg, which is connected with 12 countries. Since, this is the convergent preferential network Luxembourg can be considered as a representative European country. This network has one more interesting feature, which can not be observed in trees -- cliques. Three cliques can be distinguished in Fig.\ref{fig:PLCS_rev_str_NPT}: LU-DE-AT, LU-BE-NL, LU-NL-FR. This grouping reflects strong economical, cultural and even linguistic relationship among these countries.

\begin{figure}
 \centering
\includegraphics[scale=0.8]{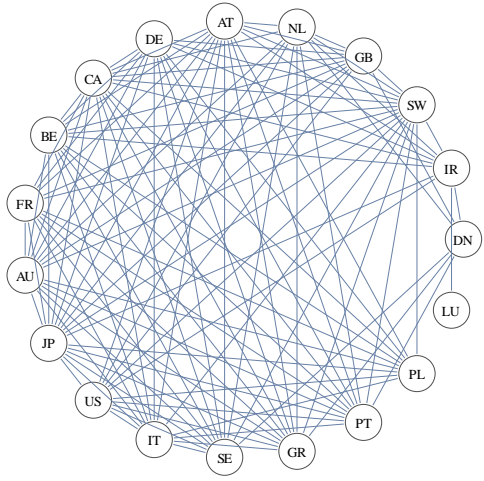} 
 \caption{Strength preferential NPT based on the PLCS correlation strength matrix  for GDP/c time series of chosen countries in years (2002, 2011). }
 \label{fig:PLCS_stab_NPT}
\end{figure}

The NPT-SP (Fig.\ref{fig:PLCS_stab_NPT}) is in some sense a complementary network to NPT-CP, since mean degree of the node is 11.8 (NPT-SP) comparing to 2.2 (NPT-CP), so the network is close the complete graph. In this case Luxembourg is the less connected country and is linked only to Ireland. While the countries with the highest number of links are: Japan (16), Sweden (16), Canada (15) and Austria (15) -- the number of links is given in brackets. It is noticeable that the mentioned countries developed countries, which are globally influential countries this agree with the main purpose of NPT-SP should point out the most influential entities.

The third type of the network -- NPT-S is constructed in order to point out the nodes, which form stable relationship with the rest of the network. In the period (2002, 2011) the nodes with the highest number of links were: Greece (12), Poland (12) and Spain (11). The common factor of these hubs is that they have relatively low GDP/c so the key factor is the main trend while the fluctuations of GDP/c are a higher order corrections in cumulative MD time series. On the other hand nodes with particularly stable relationship with the other countries might serve as a good starting point for preparing prognosis for the whole region. Unfortunately Greece and Spain are recently under a strong financial pressure which cast a shadow on the presented results.  

The NPT based on the UD of GDP/c in recent 10 years is in fact almost complete graph Fig.\ref{fig:UD_NPT}, where Belgium is connected to 18 countries, Japan is connected only to Australia therefore the remaining countries are linked to 17 nodes. 

The conclusions of ten-years period NPT analysis are also valid for twenty and thirty-years periods. NPT-CP constructed for the period (1992, 2011) point out Luxembourg as the node with highest number of links (14 countries connected). The other considered countries have significantly smaller degree (between 4 and 1). The following cliques are observed FR-NL-DE, PL-PT-IR, GR-SE-IR, LU-IT-GB, AT-LU-JP and IT-LU-JP. Comparing the set of cliques observed in twenty-years period with those seen in ten-years interval it can be noticed that the number of cliques is doubled. Moreover the clusters can be divided into two groups: high GDP/c level (FR-NL-DE, LU-IT-GB, AT-LU-JP, IT-LU-JP) and lower GDP/c level (PL-PT-IR, GR-SE-IR).

NPT-SP for the twenty year period is a network with very high number of links (mean node degree is equal 12.1). The countries with the highest number of links are Sweden (node degree: 16), USA (node degree: 15) and Portugal (node degree: 15). The less connected country is Luxembourg -- similarly to the case of NPT-SP in ten-years analysis. The stability preferential NPT based on PLCS twenty-years period analysis mark out Luxembourg and Poland as the nodes with the highest degree (17 nodes attached).  Combining the conclusions of  NPT-S and NPT-CT analysis for the twenty-years period it should be noticed that Luxembourg plays a special role for the chosen group of countries -- It is the biggest hub in convergent and stability preferential network, which indicates this country as a central points for the prognosis and economical analysis of the region.

NPT based on UD matrix for twenty year interval (1992, 2011) is very similar to the ten year analysis (Fig.\ref{fig:UD_NPT}) with the difference: Germany is the country with the maximum degree (18 links), Japan is the lowest degree node (one link) other countries have 17 links.

In the longest considered time interval (1982, 2011) the highest degree nodes of NPT-CP are Japan and Luxembourg, which are connected to 9 other nodes. In this network there is present the four nodes clique: GR-IR-SE-JP, besides that there are ten three nodes cliques: GB-JP-IT, PL-JP-IT, PT-PL-IR, CA-LU-DN, CA-US-LU, AU-US-LU, DN-NL-LU, NL-BE-LU, FR-DE-LU, IT-GR-JP. It is interesting that the Benelux countries form a clique at the thirty-years period, so this is rather long-term relationship then short term cooperation or coalition.
Analysing the strength preferential NPT-ST in the interval (1982, 2011) one can observe that Sweden is the country with the highest links attached (12 links). Considering the fact that Sweden is a highest degree node for NPT-SP for ten, twenty and thirty-years periods this country should be considered as an indicator of trend in GDP/c evolution for the most influential countries.
The last PLCS based NPT is the stability preferential NPT-S for the period (1982, 2011). In this case Poland plays the role of the main hub having 17 nodes attached. An interesting observation is that Poland belongs to the group of the main hubs for all of the considered intervals. Of course it is worth further economical analysis, but at the beginning of `80 Poland underwent economical and political transformation and is adopting its economy and law to the developed country standards, therefore its GDP/c time series are ``following'' those of developed countries. 
The UD based NPT network in the longest considered here interval is in fact difficult to analyse since there is no well established leaders or a group of the key hubs. On the network there are  nine nodes which have 17 links, five nodes with 16 nodes attached and three countries with 15 country connected. Only Luxembourg and Japan have 12 and 6  linked country respectively.

\section{Conclusions}

The analysis of correlation is a often key problem in scientific analysis particularly in economics. In the presented paper a new algorithm of correlation analysis is proposed. The algorithm is based on the power law classification scheme and the analysis of network of percolation threshold. The algorithm was applied to the analysis of correlation among GDP/c time series of 19 the most developed countries. It has been shown that PLSC is capable to recognise other then linear correlations. Particularly point out converging and diverging time series. It provides also information on the stability of correlations, which particularly important while preparing prognosis or simply justifying quality of the discovered correlation. Moreover, when applied to analysis of linear correlations  the methods is more robust to noise than the UD. Finally, the PLCS outcome was used to construct following networks: NPT-CP, NPT-SP and NPT-S. The network analysis give the opportunity to point out the entity which is 
representative for the group, with respect to analysed feature. In the case of the considered countries the representative country with respect to the strength of correlations, so the most influential one is Sweden, the most representative in the view of the similarities among GDP/c time series appeared Luxembourg and finally the most stable correlation so the best choice for the starting point for preparing prognosis was Poland.

\end{document}